\documentclass[12pt]{iopart}
\begin{document}

\title[$\Lambda{\rm CDM}$ Limit of the Generalized Chaplygin Gas]{The $\Lambda{\rm CDM}$ Limit of the Generalized Chaplygin Gas Scenario}

\author{P P Avelino\dag\ddag, L M G Beca\dag, J P M de Carvalho\S$\|$ 
and C J A P Martins\S\P$^+$}

\address{\dag\ Centro de F\'{\i}sica do Porto e Departamento de F\'{\i}sica 
da Faculdade de Ci\^encias da Universidade do Porto, Rua do Campo Alegre 687,
4169-007, Porto, Portugal}

\address{\ddag\ Astronomy Centre,
University of Sussex, Brighton BN1 9QJ, U.K.}

\address{\S\ Centro de Astrof\'{\i}sica da Universidade do Porto,
R. das Estrelas s/n, 4150-762 Porto, Portugal}

\address{$\|$\ Departamento de Matem\'atica Aplicada da Faculdade de
Ci\^encias da Universidade do Porto, Rua do Campo Alegre 687,
4169-007, Porto, Portugal}

\address{\P\ Department of Applied Mathematics and Theoretical
Physics, Centre for Mathematical Sciences, University of
Cambridge, Wilberforce Road, Cambridge CB3 0WA, U.Kingdom.}

\address{$^+$\ Institut d'Astrophysique de Paris, 98 bis Boulevard
Arago, 75014 Paris, France}

\begin{abstract}
We explicitly demonstrate that, contrary to recent claims, the dynamics of 
a generalized Chaplygin gas model with an equation of state $p=-C$ (where 
$C$ is a positive constant) is equivalent to that of a standard $\Lambda$CDM 
model to first order in the metric perturbations. We further argue that the 
analogy between the two models goes well beyond linear theory and conclude 
that they cannot be distinguished based on gravity alone.
\end{abstract}

\ead{C.J.A.P.Martins@damtp.cam.ac.uk}
\maketitle

\section{Introduction}

Observational evidence strongly suggests that we live in 
a (nearly) flat Universe which has recently entered an accelerating phase 
\cite{Perlmutter,Riess,Tonry,wmap}. In the context of general relativity 
such a period of accelerated expansion must be induced by an exotic 
`dark energy' component violating the strong energy 
condition \cite{lambda,picon,Wang,solid,Bagla}, though this is not
necessarily so in the context of more general models 
(see for example \cite{brane}). There is also strong evidence that most 
of the matter in the Universe, an essential ingredient for structure 
formation, is in a non-baryonic form.

One can therefore ask if it is possible to have some component of the
energy budget of the universe which simultaneously accounts for both
the dark energy and the dark matter, or if two different components
are inevitable. An interesting toy model candidate for the unification of dark 
matter and dark energy is a perfect fluid with an exotic equation of 
state known as the (generalized) Chaplygin gas, for which 
\begin{equation}
p=-\frac{C}{\rho^\alpha}\,.
\label{eqnstate}
\end{equation}
Here $\rho$ is the density, $p$ is the 
pressure, $C$ is a positive constant and $0\leq\alpha\leq1$. 
Both the best-motivated Chaplygin gas (which has $\alpha=1$)
\cite{Kamen,Bilic} and simple generalizations thereof \cite{Bento} have 
recently attracted considerable attention at this toy model level.
In fact a connection between string theory and the original Chaplygin gas 
has also been claimed (see for example \cite{Hassaine} and references therein).

In the $\alpha=0$ case it is straightforward to show that 
the background equations for the Chaplygin gas model are identical 
to those of the familiar $\Lambda$CDM scenario. However, there has been 
a recent claim \cite{equivalence} that this similarity breaks down 
at first order in the metric perturbations.

In fact, we have recently shown \cite{nonlinear} that for $\alpha >0$ the
computation of precise predictions in the context of this type of models 
does need to take into account the non-linear behavior since the background
equations will cease  to be valid at late times even on large cosmological
scales. In broad terms, this is related to the fact that in general
\begin{equation}\label{inequality}
\langle p \rangle  \equiv 
- \langle C / \rho^\alpha  \rangle \neq - C / \langle \rho \rangle^\alpha\, ,
\end{equation} 
where $\langle \ \rangle$ represents a spatial average.
This means that all previous 
predictions \cite{Fabris1,Fabris,Avelino,Dev,Gorini,Makler,Alcaniz,Finelli,
Sandvik,Bean,Bento1,Beca,Amendola} relying on the background evolution or 
linear evolution of density perturbations must be
re-evaluated. However, this is only a problem if $\alpha \neq 0$.

In this paper we explicitly show that, contrary to the claim in
\cite{equivalence}, the evolution of linear density perturbations in 
the context of a Chaplygin gas model with  
$\alpha=0$ is identical to that of a standard $\Lambda$CDM model. We further 
conclude that these models are in fact indistinguishable as far as gravity is 
concerned. 
 
\section{The background dynamics}

The dynamics of a flat homogeneous and isotropic Friedmann-Robertson-Walker 
(FRW) universe is fully described by 
\begin{eqnarray}
\label{eq1}
\left(\frac{\dot a}{a}\right)^2=\frac{8 \pi G \rho a^2}{3} \, , \\
\label{eq2}
\left(\frac{\ddot a}{a}\right)-\left(\frac{\dot a}{a}\right)^2=
-\frac{4 \pi G (\rho+3p) a^2}{3} \, ,
\end{eqnarray}
if the equation of state $p = p(\rho)$ is provided. Here a dot represents 
a conformal time derivative ($d/d\eta$), $a$ is the scale 
factor, $\rho$ is the energy density and $p$ is the pressure. 
It is possible to show using Eqs. (\ref{eq1}) and (\ref{eq2}) that 
\begin{equation}
\label{eq2a}
{\dot \rho} + 3\frac{\dot a}{a}(\rho+p)=0 \, .
\end{equation}

Now, let us consider two possible scenarios: 
\begin{itemize}
\item Model I:  a generalized Chaplygin gas model with a density 
$\rho_{\rm{I}}$, and pressure $p_{\rm{I}}=-C$
\item Model II: a $\Lambda$CDM scenario with pressureless cold dark 
matter with density $\rho_m$, and a cosmological constant with density 
$\rho_\Lambda$, and pressure $p_\Lambda=-\rho_\Lambda=-C$
\end{itemize}
where $C$ is a positive constant.
It is trivial to verify that the background equations (and hence the
dynamics) for models I and II are identical when one identifies
the total densities and pressures in both cases, that is 
\begin{equation}
\label{ident1}
\rho_{\rm{I}} = \rho_{\rm{II}} = \rho_m + \rho_\Lambda\, ,
\end{equation}
\begin{equation}
\label{ident2}
p_{\rm{I}} = p_{\rm{II}} = p_\Lambda = -C \, .
\end{equation}
We can therefore physically interpret $C$ in the Chaplygin gas case 
as the equivalent vacuum energy density of the $\Lambda$CDM model.

\section{Growth of Perturbations}

For a general case of $n$ gravitationally interacting fluids, the linear
evolution of perturbations in the synchronous gauge is
given by \cite{Shoba}:
\begin{eqnarray}
\label{eq3}
{\ddot h} + {\cal H}{\dot h} + 3 {\cal H}^2\sum\nolimits_i {(1 + 3v_i^2 })\Omega_i\delta_i=0 \\
\label{eq4}
{\dot \delta}_i +(1+\omega_i)(\theta_i + {\dot h}/2)+ 3{\cal H}(v_i^2-\omega_i)\delta_i=0 \\
\label{eq5}
{\dot \theta_i}+{\cal H}(1-3v_i^2)\theta_i+ \frac{v_i^2}{1+\omega_i}\nabla^2\delta_i=0
\end{eqnarray}
where $h$ is the trace of the perturbation to FRW metric, ${\cal H} \equiv {\dot a}/a$, 
$\delta_i$ is the density contrast of the $i$th-fluid obeying 
$p_i=\omega_i\rho_i$ with an adiabatic sound speed $v_i$ and an element 
velocity divergence  $\theta_i$. Note that Eqs. (\ref{eq4}) and (\ref{eq5}) 
apply for all $i=1,\ldots,n$. 

In models I and II the sound speed is identically zero so 
that Eq. (\ref{eq5}) reduces to $\theta_i=0$ at all times in the initially unperturbed synchronous gauge. Using the fact that 
${\dot w}_{{\rm{I}}}/w_{{\rm{I}}}= 3 {\cal H}(1+w_{{\rm{I}}})$ and 
\begin{equation}
\label{eq6}
{\dot h} = \frac{-2 {\dot \delta}_{{\rm{I}}} +6\omega_{{\rm{I}}} {\cal H} \delta_{{\rm{I}}}}{1+\omega_{{\rm{I}}}}=-2\frac{d}{d\eta} \left(\frac{\delta_{\rm{I}}}{1+w_{\rm{I}}}\right),
\end{equation}
it is straightforward to show that the evolution of density perturbations in model I is given by
\begin{equation}
\label{eq7}
{\ddot \delta_*} + {\cal H}{\dot \delta_*} - \frac{3}{2}{\cal H}^2(1+\omega_{{\rm{I}}})\delta_*=0,
\end{equation}
where $\delta_* = \delta_{{\rm{I}}}/(1+w_{{\rm{I}}})$. Let us now consider the evolution of matter perturbations, $\delta_m$, in the context of model II. Using 
equation (\ref{eq3}) plus the fact that $\dot h=-2 \dot \delta_m$ and the relation $\Omega_m = \rho_m/(\rho_m + \rho_\Lambda) =  1+w_{\rm{II}}$ it is easy to show that
\begin{equation}
\label{eq8}
{\ddot \delta_m} + {\cal H}{\dot \delta_m} - \frac{3}{2}{\cal H}^2(1 +w_{\rm{II}}) \delta_m=0.
\end{equation} 
It is now immediate to see that the evolution of $\delta_{{\rm{I}}}$ and 
$\delta_{\rm{II}} \equiv \delta \rho_{\rm{II}} / \rho_{\rm{II}} = 
\delta_m /(1+\rho_\Lambda/\rho_m)=\delta_m (1+ w_{\rm{II}})$ is identical 
if we identify $w_{\rm{II}}$ with $w_{\rm{I}}$. It is straightforward to show 
from Eq. (\ref{eq4}) that $\delta_{\rm I,II}$ has an asymptotic behavior $\propto a^{-3}$ at late times.
In reference \cite{equivalence} the authors wrongly concluded that the 
Chaplygin gas model with $\alpha=0$ and $\Lambda$CDM would differ 
in their first order  evolution on the 
basis of their different evolutions in $\delta_m$ and $\delta_{\rm{I}}$. 
As we have explicitly shown above, these are not the right 
variables to compare. 

In fact, given that the evolution of the density perturbations in 
Fourier space for models I and II is independent of the wave-number 
$k$, the fact that the models are equivalent to zeroth order in the 
metric perturbations necessarily implies their equivalence to first 
order. Note that a perturbation with an infinite wavelength ($k=0$) 
is uniform and can be studied using the equations for the background 
evolution.

\section{Beyond linear theory}

It is also possible to show that the models are in fact equivalent 
beyond first order in the metric perturbations. From the point of
view of gravity, the models are equivalent 
if, given the same initial conditions, the evolution of the metric and 
the energy-momentum tensor of the perfect fluid driving the expansion 
of the Universe,
\begin{equation}
\label{eq9}
T^{\alpha \beta}=(\rho+p)u^{\alpha} u^{\beta} + p g^{\alpha \beta}\, ,
\end{equation} 
is identical for both models. To specify the evolution of the energy-momentum 
tensor one 
needs to know the evolution of $\rho$ and $3$ of the components of the 
4-velocity $u^{\alpha}$. Note that the constant pressure is specified and 
that the 4th component of the 4-velocity can be determined from the other $3$ 
using the condition $u^{\alpha} u_{\alpha} = -1$. 

On the other hand, given initial conditions at the time $t_i$ 
for the components of the metric $g^{\alpha \beta}$, 
and their first derivatives it is possible to use the Einstein equations to 
compute the second derivatives of the metric with respect to time everywhere, 
and use this 
to calculate the new values of the metric components and their first 
derivatives at a 
subsequent time $t_i+dt$. 

Also, the conservation of the energy-momentum 
tensor (${T^{\alpha \beta}}_{;\alpha}=0$) gives us $4$ equations which 
relate the components of the energy-momentum tensor at the instants 
$t_i+dt$ and $t_i$. Given that the evolution of the energy-momentum tensor 
depends on $4$ variables (other than the metric) it is completely 
determined by the 4 equations which need to be satisfied in order for the 
energy-momentum tensor conservation to hold.  

Hence, we conclude that if the initial conditions are the same for both models 
(I and II) the dynamics will be identical so that the models can not be 
distinguished based on gravity alone.

\section{Conclusions}

We have shown that gravity alone cannot distinguish between a generalized 
Chaplygin gas model with $\alpha=0$ and a standard $\Lambda$CDM scenario.
Given similar initial conditions, their dynamics will be similar so 
that the cosmological predictions of both models are identical. 
Given that both of them are, to a certain extent, 
toy models with a somewhat tenuous motivation from fundamental physics, 
this is probably as far as they can meaningfully be compared.

In any case, our present results, together with those of \cite{nonlinear}
show that the equation of state \ref{eqnstate} with $0\leq\alpha\leq1$
basically describes a one parameter family of toy models interpolating
from $\Lambda$CDM to the original Chaplygin gas model. As such they can be 
useful for studying cosmological constraints on unified dark matter models,
though one must keep in mind that as soon as $\alpha\neq0$ the analysis
becomes quite subtle.

\ack
We are grateful to Miguel Costa and Carlos Herdeiro for useful discussions.
P.A. was partially funded by FCT (Portugal), under grant SFRH/BSAB/331/2002.
C.M. is funded by FCT (Portugal), under grant FMRH/BPD/1600/2000.
Additional support came from FCT under contract CERN/POCTI/49507/2002.

\end{document}